\documentclass[12pt]{article}
\usepackage{amsfonts}
\usepackage{amssymb}
\usepackage{amsbsy}
\usepackage{mathrsfs}
\usepackage{amsmath}
\usepackage{amsthm}
\usepackage{graphicx}    
\usepackage{rotating}    
\usepackage{epsfig}
\usepackage{color}       

\usepackage{bm}              
\usepackage{multirow}        

\DeclareFontFamily{OT1}{pzc}{}
\DeclareFontShape{OT1}{pzc}{m}{it}{<-> s * [1.200] pzcmi7t}{}
\DeclareMathAlphabet{\mathpzc}{OT1}{pzc}{m}{it}
\input epsf.sty

\newlength{\TZ}
\newcommand{\BEQ}{\begin{equation}}     
\newcommand{\BEA}{\begin{eqnarray}}
\newcommand{\BD}{\begin{displaymath}}
\newcommand{\EEQ}{\end{equation}}       
\newcommand{\EEA}{\end{eqnarray}}
\newcommand{\ED}{\end{displaymath}}
\newcommand{\D}{{\rm d}}                
\newcommand{\II}{{\rm i}}               
\newcommand{\demi}{\frac{1}{2}}         
\newcommand{\wit}[1]{\widetilde{#1}}    
\newcommand{\lap}[1]{\overline{#1}}     
\newcommand{\dlap}[1]{\overline{\overline{#1}}}    

\newcommand{\Vek}[1]{\boldsymbol{#1}}    

\newcommand{\fns}{\footnotesize}  

\catcode`\@=11
\def\numberbysection{\@addtoreset{equation}{section}
        \def\theequation{\thesection.\arabic{equation}}}
\numberbysection

\begin{document}

\begin{titlepage}

\vskip 1.5 cm
\begin{center}
{\LARGE \bf Quantum dynamics far from equilibrium: \\[0.05truecm] a case study in the spherical model\footnote{Conference proceedings of LT14, 20$^{\rm th}$-26$^{\rm th}$ of June 2021, Sofia (Bulgarie)}}
\end{center}

\vskip 2.0 cm
\centerline{{\bf Malte Henkel}$^{a,b}$}
\vskip 0.5 cm
\centerline{$^a$Laboratoire de Physique et Chimie Th\'eoriques (CNRS UMR 7019),}
\centerline{Universit\'e de Lorraine Nancy, B.P. 70239, F -- 54506 Vand{\oe}uvre l\`es Nancy Cedex, France}
\vspace{0.5cm}
\centerline{$^b$Centro de F\'{i}sica Te\'{o}rica e Computacional, Universidade de Lisboa,}
\centerline{Campo Grande, P--1749-016 Lisboa, Portugal}

\begin{abstract}
The application of quantum Langevin equations for the study of non-equilibrium relaxations is illustrated in the exactly solved quantum
spherical model. Tutorial sections on the physical background of non-markovian quantum noise, the spherical model quantum phase transition, the long-time limit of
the quantum Langevin equation of the spherical model and physical ageing are 
followed by a brief review of the solution of the non-markovian time-dependent spherical constraint and about the consequences for quantum ageing 
at zero temperature, after a quantum quench. 
\end{abstract}
\end{titlepage}

\setcounter{footnote}{0}

\section{What is a quantum Langevin equation~?}

The description of quantum-mechanical many-body problems far from equilibrium 
presents conceptual difficulties which go beyond those present in classical
systems \cite{Breu02,Engl02,Cugl03,Gard04,Taeu14,Cald14,Oliv20,Weis21}. Physically, one distinguishes {\em closed} systems, which are 
isolated and {\em open} systems, which are coupled to one or several external baths. For closed systems, the Heisenberg equations
of motion are a convenient starting point. 

For open quantum systems, a large variety of theoretical descriptions has been considered. 
Here, we shall concentrate on quantum Langevin equations, where the `noises' must be chosen as to (i) maintain the quantum coherence
of the system and (ii) to describe the interaction with the external baths.  
We shall begin with a tutorial for the formulation of dissipative quantum dynamics of open system and shall use the quantum spherical model for a
case study.\footnote{For studies of the quantum dynamics in closed spherical models, see \cite{Chan13,Mara15,Cugl17,Cugl18,Barb19,Hali21}.} 
In this section~1, we recall the generic formulation of quantum Langevin equations, in section~2 we give a brief introduction to
the quantum spherical model and its quantum phase transition, 
in section~3 the quantum dynamics of this model is formulated in a way to facilitate the extraction of the long-time
behaviour of physical observables and in section~4, we recall the main ingredients to describe the physical ageing expected in relaxational
dynamics after a quantum quench. The later sections review the results of a detailed analysis in the quantum spherical model,
at temperature $T=0$. 
Section~5 discusses the solution of the non-linear integral equation derived from spherical constraint. Section~6 and~7 review the
results after quenches to either the disordered phase or else onto the critical point or into the ordered phase. We conclude in section~8.

Inspired by a proposal of Bedeaux and Mazur \cite{Bede01,Bede02}, we consider quantum Langevin equations in the following form,
for simplicity formulated for a single quantum variable $s$ and its canonically conjugate momentum $p$, 
\BEQ \label{1.1}
\partial_t s = \frac{\II}{\hbar} \bigl[ H, s \bigr] + \eta^{(s)} \;\; ~~;~~ \;\; 
\partial_t p = \frac{\II}{\hbar} \bigl[ H, p \bigr] -\gamma p + \eta^{(p)}
\EEQ
where $H$ is the hamiltonian of the system and the damping constant $\gamma>0$ describes the dissipative part of the dynamics. The moments of
the noise operators $\eta^{(s)},\eta^{(p)}$ must be specified as to maintain the required quantum properties of the dynamics. 

\begin{figure}[tb]
\includegraphics[width=.5\hsize]{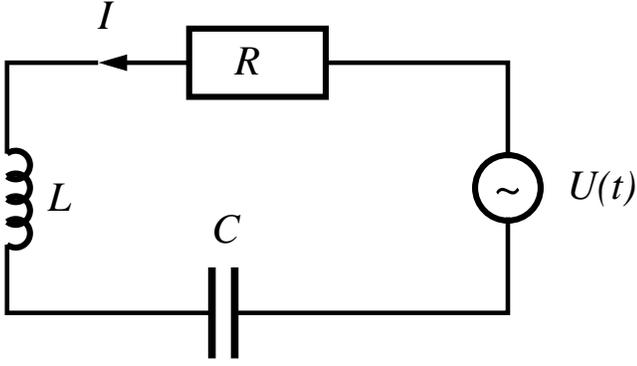} ~~~
\caption[fig1]{Schematic LRC circuit. \label{fig1}}
\end{figure}

That the noise structure in (\ref{1.1}) is quite natural can be seen from the example of a LRC electric circuit \cite{Becker78}, see fig.~\ref{fig1}.  
According to Kirchhoff, one has $U(t)=U_R+U_L+U_C$ and furthermore $U_R = R I$ and $U_L = L \dot{I}$, where $I=I(t)$ is the
current and $R$, $L$ are the resistance and the inductivity, respectively. Their combined noises are modeled by setting $U(t)=L \eta_U$. In addition,
the (noisy) voltage fluctuations at the capacity $C$ are described by $\dot{U}_C=\frac{1}{C}I+\eta_I$. This leads to 
\BEQ \label{1.2}
\partial_t U_C = \frac{1}{C} I + \eta_I \;\; ~~;~~ \;\; 
\partial_t I = - \frac{1}{L} U_C -\frac{R}{L} I  + \eta_U
\EEQ
and with the correspondences $s \leftrightarrow U_C$, $p \leftrightarrow I$ and
$\eta^{(s)} \leftrightarrow \eta_I$, $\eta^{(p)} \leftrightarrow \eta_U$, (\ref{1.2}) is identified with (\ref{1.1}), 
if $H$ describes a harmonic oscillator. Apparently the noises $\eta_U,\eta_I$ describe the `rough' fluctuations around the
smooth averages $I=I(t)$ and $U_C=U_C(t)$. This example also suggests that in the context of nano-electronics, quantum noise
effects might become of relevance. 

Returning to (\ref{1.1}), it remains to specify the noise correlators. For definiteness, it is assumed that any deterministic term
is included into the hamiltonian $H$, so that $\bigl\langle \eta^{(s)}\bigr\rangle =  \bigl\langle \eta^{(p)}\bigr\rangle = 0$. The non-vanishing
second moments are, at temperature $T>0$  
\BEQ \label{1.3} 
\bigl\langle \bigl\{ \eta^{(s)}(t), \eta^{(p)}(t')\bigr\} \bigr\rangle = \gamma T \coth\left(\frac{\pi}{\hbar}T \bigl(t-t'\bigr) \right) 
\;\; ~;~ \;\;
\bigl\langle \bigl[ \eta^{(s)}(t), \eta^{(p)}(t')\bigr] \bigr\rangle = \II\hbar \gamma\,\delta(t-t')
\EEQ
Clearly, quantum noise is explicitly non-markovian.\footnote{In the classical limit $\hbar\to 0$, markovian white-noise
correlators are recovered from (\ref{1.3}) \cite{Arau19}.} Anti-commutators $\{.,.\}$ and commutators $[.,.]$ were used. 
Eq.~(\ref{1.3}) can be derived in two distinct ways: 
\begin{enumerate}
\item The classical approach of Ford, Kac and Mazur \cite{Ford65,Ford87,Ford88} considers explicitly the coupling of the system 
to an external bath, with the total hamiltonian $H_{\rm tot}=H+H_{\rm int}+H_{\rm bath}$. 
The composite object described by $H_{\rm tot}$ is considered as a closed system. 
Using for the bath hamiltonian $H_{\rm bath}$ a large ensemble of harmonic oscillators,
along with a bi-linear coupling $H_{\rm int}$ to the system, the Heisenberg equations of the bath degrees of freedom can be formally solved. 
An average over the initial positions and momenta of the bath then leads to (\ref{1.3}). Herein, the system is a single degree of freedom $s$ in some
external potential $V=V(s)$. 
\item A phenomenological derivation \cite{Arau19} considers the `desirable' physical properties of dissipative quantum dynamics which
any choice of the quantum noises should keep. These are
\begin{enumerate}
\item canonical equal-time commutator $\bigl\langle [s(t),p(t)]\bigr\rangle = \II\hbar$. 
\item the Kubo formula of linear response theory \cite{Pari88,Cugl03}. 
\item the virial theorem (for selecting equilibrium stationary states) \cite{Fock30,Schn08}. 
\item the quantum fluctuation-dissipation theorem ({\sc qfdt}) (to distinguish quantum and classical equilibrium states) \cite{Ford17,Hang05}. 
\end{enumerate}
While the {\sc qfdt} is habitually formulated in frequency space, for any temperature $T>0$ a mathematically equivalent statement is the
Kubo-Martin-Schwinger relation \cite{Pari88,Cugl03,Arau19}
\BEQ \label{1.4}
C\left(t-t'+\frac{\II\hbar}{2T}\right) - C\left(t-t'-\frac{\II\hbar}{2T}\right) = \frac{\hbar}{2\II} 
\left[ R\left(t-t'+\frac{\II\hbar}{2T}\right) + R\left(t-t'-\frac{\II\hbar}{2T}\right) \right]
\EEQ
where $C(t-t') = \demi\bigl\langle\bigl\{ s(t), s(t')\bigr\}\bigr\rangle$ is the stationary correlator and
$R(t-t')=\left. \frac{\delta \bigl\langle s(t)\bigr\rangle}{\delta h(t')}\right|_{h=0}$ 
is the stationary linear response with respect to the conjugate field $h$. The {\sc qfdt} follows from a Fourier transformation of (\ref{1.4}) 
with respect to $t-t'$. If the chosen system is a harmonic oscillator, its dynamics can be formally solved which provides a relation between
the noise correlators and the four physical criteria raised above. Postulating that the noise correlators should not contain any system-specific 
parameter, eq.~(\ref{1.3}) follows \cite{Arau19}. 
\end{enumerate} 
That these distinct approaches lead to the same noise correlators (\ref{1.3}) also clarifies important physical properties of this choice of 
dynamics. The only physical parameters are the dissipation constant $\gamma$, the bath temperature $T$ and Planck's constant $\hbar$. 
The validity of (\ref{1.3}) does not depend on the implicit auxiliary assumptions contained in either approach. 
Eqs.~(\ref{1.1},\ref{1.3}) are the {\em quantum Langevin equations}, to be used in what follows. 

\begin{figure}[t]
\includegraphics[width=.95\hsize]{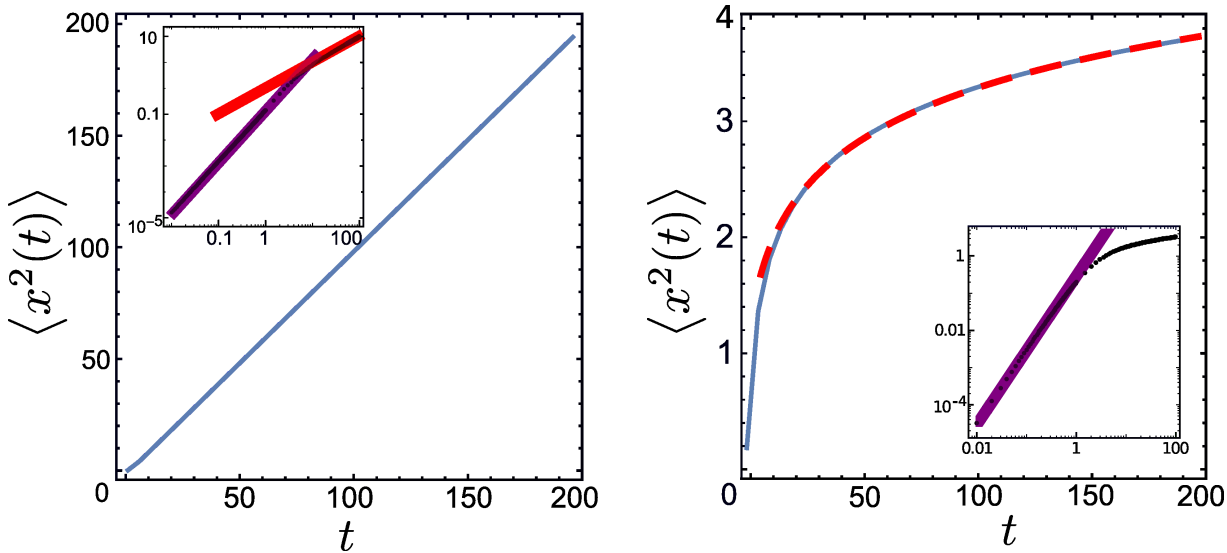} 
\caption[fig2]{Variance $\bigl\langle x^2(t)\bigr\rangle$ of a brownian particle for classical white noise with $T>0$ (left panel) and
for quantum noise (\ref{1.3}) at $T=0$ (right panel). Insets: cross-over from initial ballistic motion. 
\label{fig2}}
\end{figure}
The qualitative differences between classical and quantum noises can be illustrated through the motion of a free $1D$ brownian particle. 
Figure~\ref{fig2} \cite{Arau19} compares the variance $\bigl\langle x^2(t)\bigr\rangle\sim t$ of the position $x(t)$ for classical white noise
at finite temperature $T>0$ (left panel) with the quantum result $\bigl\langle x^2(t)\bigr\rangle\sim \ln t$ obtained at $T=0$ (right panel). 
Consequently, quantum diffusion is `more weak' than classical diffusion since it needs considerably more time to homegenise a system.

Excellent reviews of dissipative quantum dynamics include \cite{Gard04,Cald14,Weis21}.

\section{What is the quantum spherical model~?}

The {\em spherical model} is a simple, yet non-trivial, and exactly solvable model for the study of phase transitions \cite{Berl52}. 
Its classical version is defined in terms of a continuous spin variable $s_{\Vek{n}}\in\mathbb{R}$ attached to the sites 
$\Vek{n}\in\mathscr{L}$ of a $d$-dimensional hyper-cubic lattice $\mathscr{L}\subset\mathbb{Z}^d$. In its most simple variant, 
one uses nearest-neighbour interactions ${\cal H}=-\sum_{(\Vek{n},\Vek{m})} s_{\Vek{n}} s_{\Vek{m}} +\frac{\mu}{2} \sum_{\Vek{n}} s_{\Vek{n}}^2$,
where the Lagrange multiplier $\mu$ is fixed from the (mean) {\em spherical constraint} \cite{Berl52,Lewi52}
\BEQ \label{2.1}
\left\langle \sum_{\Vek{n}\in\mathscr{L}} s_{\Vek{n}}^2 \right\rangle = \mathscr{N}
\EEQ
where $\mathscr{N}=|\mathscr{L}|$ is the number of sites of the lattice $\mathscr{L}$ and 
$\bigl\langle \cdot\bigr\rangle$ denotes the thermodynamic average. This condition was originally motivated by a comparison
with the Ising model, with discrete `Ising spins' $s_{\Vek{n}}=\pm 1$ and which naturally obey (\ref{2.1}) \cite{Berl52}. The spherical model
is solvable since in Fourier space, the degrees of freedom decouple, but some interactions do remain because of the constraint (\ref{2.1}). 
At equilibrium, the spherical model has a critical point $T_c>0$ for any spatial dimension $d>2$ and the universality class of the model
is distinct from mean-field theory if $2<d<4$. The values of the critical exponents are different from those found in the Ising model, e.g. \cite{Taeu14}. 

\begin{figure}
\includegraphics[width=.6\hsize]{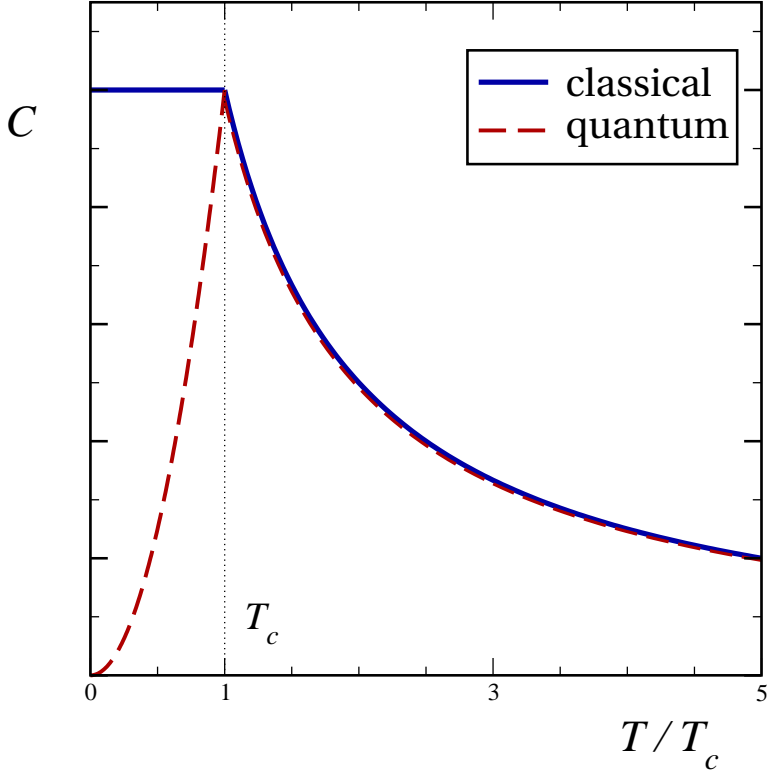} 
\caption[fig3]{Specific heat $C$ (schematic) in the classical and quantum spherical models. \label{fig3}}
\end{figure}

A serious physical short-coming of the classical spherical model is its low-temperature behaviour \cite{Berl52}, see fig.~\ref{fig3}
for dimensions $2<d<4$.  The cusp of the specific heat $C$ at $T\simeq T_c$  (rather than a jump) is a manifestation of non-mean-field criticality. 
But for all temperatures $T\leq T_c$, $C$ is constant! This behaviour violates the third fundamental theorem of thermodynamics \cite{Becker78}, 
which requests that $C(T)\to 0$ as $T\to 0$. 

\begin{figure}
\includegraphics[width=1.0\hsize]{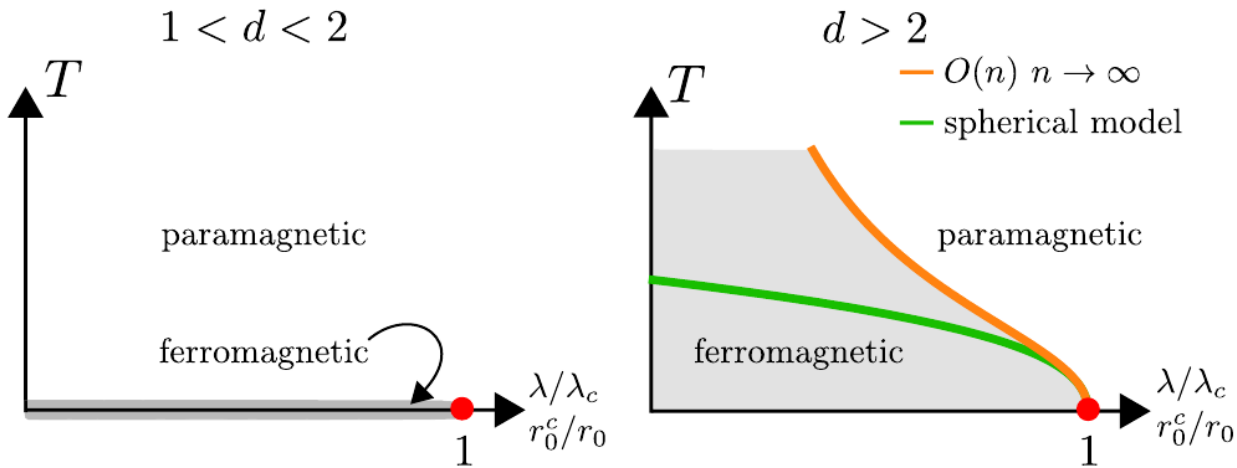} 
\caption[fig4]{Quantum and classical equilibrium phase transitions in the spherical model. \label{fig4}}
\end{figure}

The quantum spherical model corrects this deficiency. In terms of spin operators $s_{\Vek{n}}$ and conjugate momenta
$p_{\Vek{m}}$, which obey $\bigl[s_{\Vek{n}},p_{\Vek{m}}\bigr]=\II\hbar$, let \cite{Ober72,Henk84a,Niew95,Vojt96}
\BEQ \label{2.2}
H = \frac{1}{2} \sum_{\Vek{n}\in\mathscr{L}} \left[ p_{\Vek{n}}^2 + \bigl(r+d\bigr) s_{\Vek{n}}^2 - 
\sum_{(\Vek{n},\Vek{m})} s_{\Vek{n}} s_{\Vek{m}} \right] \;\; ~~;~~ \;\;
\sum_{\Vek{n}\in\mathscr{L}}  \bigl\langle s_{\Vek{n}}^2\bigr\rangle = \frac{\mathscr{N}}{\lambda}
\EEQ
(after several re-scalings) and the Lagrange multiplier is now $r := \mu\lambda -d$. The quantum hamiltonian $H$ arises from its classical
counterpart $\cal H$ by adding a kinetic energy term $\sim \sum_{\Vek{n}}p_{\Vek{n}}^2$ and $\lambda$ controls the relative importance of this term. 
Schematically, the behaviour of the quantum model
is illustrated in figure~\ref{fig3}. Indeed, for $d>2$ there exists a critical temperature $T_c>0$ such that the high-temperature
behaviour of the quantum model is analogous to the classical one. Especially, the type of singularity of $C$ around $T\simeq T_c$ is precisely
the same \cite{Niew95,Oliv06}. On the other hand, for $T<T_c$, the behaviour of the quantum model is different from the one of its classical
variant and one finds $C(T)\to 0$ as $T\to 0$, as expected from the third fundamental theorem of thermodynamics.\footnote{The same universality class is
also obtained for the $n\to\infty$ limit of the O($n$) vector model. It is well-known folklore that this holds both for classical and quantum equilibrium, as well
as for classical dynamics. It can also be shown that this universality extends to quantum dynamics \cite{Wald21}, but for the sake of brevity 
we shall not discuss this any further here.
In fig.\ref{fig4} the phase diagramme is also indicated for the O($\infty$)-model. } 

In addition, at zero temperature $T=0$, the quantum spherical model has a {\em quantum phase transition}, where $\lambda$ acts as thermodynamic 
parameter \cite{Henk84a,Niew95,Vojt96,Oliv06,Wald15}. The schematic phase diagrammes at equilibrium are shown in fig.~\ref{fig4} \cite{Wald21}. 
For dimensions $1<d<2$, there is only a quantum phase transition at $T=0$ and at $\lambda=\lambda_c>0$. 
In the ordered ferromagnetic phase, the system acquires a non-vanishing spontaneous magnetisation, but in the
disordered paramagnetic phase, it remains non-magnetic. On the other hand, 
for $d>2$, there exists not only a quantum phase transition at $T=0$ and $\lambda=\lambda_c>0$, 
but also a classical finite-temperature phase transition at some $T_c=T_c(\lambda)>0$ if $\lambda<\lambda_c$. 
The universal properties of the quantum phase transition, notably the values of the critical exponents around $\lambda\simeq\lambda_c$, 
of the $d$-dimensional quantum model are the same as the ones of the $(d+1)$-dimensional classical model around $T\simeq T_c(\lambda)$
\cite{Kogu79,Henk84a,Niew95,Vojt96,Oliv06,Sach11,Wald15}. 

The existence and nature of phase transitions are often expressed via upper and lower critical dimensions. 
The {\em lower critical dimension} $d_{\ell}$ is defined such that for $d<d_{\ell}$, no phase transition exists. The {\em upper critical dimension}
$d_u$ is defined such that for $d>d_u$, the critical behaviour of the model is identical to the one of mean-field theory. 
At equilibrium, one has for the (short-ranged) spherical model
\BEQ \label{2.3} 
\left\{ \begin{array}{lcll} d_{\ell} = 2 & ; & d_u = 4 & \mbox{\rm ~~classical} \\
                            d_{\ell} = 1 & ; & d_u = 3 & \mbox{\rm ~~quantum} \end{array} \right.
\EEQ
These values will be needed below when discussing the relaxational dynamics of the quantum spherical model.

\section{How to formulate quantum dynamics of the spherical model~?}

In order to extract the long-time and large-distance properties of the relaxational dynamics of the spherical model, 
we first carry out a continuum limit\footnote{The form (\ref{2.2}) of $H$ is chosen to facilitate taking this limit.}  
and let $s_{\Vek{n}} \mapsto \phi(t,\Vek{x})$ and $p_{\Vek{n}}\mapsto \pi(t,\Vek{x})$. 
The spherical model degrees of freedom decouple in Fourier space; hence we set 
\BEQ
\phi_{\Vek{k}}(t) := \int_{\mathbb{R}^d} \!\D \Vek{x}\: \phi(t,\Vek{x})\, e^{-\II \Vek{k}\cdot\Vek{x}} 
\EEQ
such that the quantum Langevin equations (\ref{1.1},\ref{1.3}) become in Fourier space
\BEQ \label{3.2}
\partial_t \phi_{\Vek{k}} = \frac{\II}{\hbar} \bigl[ H, \phi_{\Vek{k}}\bigr] +\eta_{\Vek{k}}^{(\phi)} \;\; ; \;\; 
\partial_t \pi_{\Vek{k}} = \frac{\II}{\hbar} \bigl[ H, \pi_{\Vek{k}}\bigr] -\gamma \pi_{\Vek{k}} +\eta_{\Vek{k}}^{(\pi)}
\EEQ
together with the non-vanishing moments 
\begin{subequations} \label{3.3} \begin{align}
\bigl\langle \bigl\{ \eta_{\Vek{k}}^{(\phi)}(t),\eta_{\Vek{k}'}^{(\pi)}(t') \bigr\}\bigr\rangle 
&= \gamma T \coth\left(\frac{\pi T}{\hbar} (t-t')\right) \delta(\Vek{k}+\Vek{k}') \\
\bigl\langle \bigl[ \eta_{\Vek{k}}^{(\phi)}(t),\eta_{\Vek{k}'}^{(\pi)}(t') \bigr]\bigr\rangle 
&= \II\hbar \gamma \delta(t-t')\, \delta(\Vek{k}+\Vek{k}')
\end{align}\end{subequations}
These describe a set of quantum harmonic oscillators which are only coupled through the spherical constraint (\ref{2.2}). 
Since the Lagrange multiplier $r=r(t)$ is now time-dependent, the explicit solution of eqs.~(\ref{3.2},\ref{3.3}) becomes very cumbersome. 

But since we shall be mainly interested in the long-time dynamics of the model, we shall project onto this long-time regime 
by carrying out a scaling transformation. For notational simplicity, we give the procedure for a single
degree of freedom \cite{Arau19,Wald21} and shall write for a moment $\phi_{\Vek{k}}(t)\mapsto \phi(t)$ and so on. Consider 
\BEQ
\tilde{t} := \lambda t \;\; ; \;\; \phi(t) = \lambda \wit{\phi}(\tilde{t}\,) \;\; ; \;\; r(t) = r(\tilde{t}) 
\EEQ
Then the quantum Langevin equation becomes
\begin{subequations}\begin{align} 
\lambda^2 \partial_{\tilde{t}}^2 \wit{\phi}(\tilde{t}\,) &= -r(\tilde{t}) \wit{\phi}(\tilde{t}\,) 
- \tilde{\gamma} \partial_{\tilde{t}}\wit{\phi}(\tilde{t}\,)  + \wit{\xi}(\tilde{t}\,) \label{3.5a} \\
\wit{\xi}(\tilde{t}\,) &:= \wit{\eta}^{(\pi)}(\tilde{t}\,) + \tilde{\gamma} \lambda^{-2} \wit{\eta}^{(\phi)}(\tilde{t}\,)
+\partial_{\tilde{t}}\wit{\eta}^{(\phi)}(\tilde{t}\,)
\end{align}\end{subequations}
where we set $\tilde{\gamma} := \lambda \gamma$. In the long-time scaling limit
\BEQ \label{3.4}
t\to \infty \;\; , \;\; \lambda\to 0 \;\; , \;\; \mbox{\rm such that~ $\tilde{t}=\lambda t$ ~is kept fixed}
\EEQ
the quantum equation of motion (\ref{3.5a}) reduces to an over-damped Langevin equation 
\begin{subequations} \label{3.6}
\BEQ
\tilde{\gamma} \partial_{\tilde{t}}\wit{\phi}(\tilde{t}\,)  = -r(\tilde{t}) \wit{\phi}(\tilde{t}\,)  + \wit{\xi}(\tilde{t}\,) 
\EEQ
with the noise correlators 
\BEQ \label{3.6b}
\bigl\langle \bigl\{ \wit{\xi}(\tilde{t}\,), \wit{\xi}(\tilde{t}'\,)\bigr\}\bigr\rangle = \frac{\hbar\tilde{\gamma}}{\pi}
I\left( \frac{\hbar}{2\tilde{T}}, \tilde{t}-\tilde{t}'\right) \;\; ; \;\;
\bigl\langle \bigl[ \wit{\xi}(\tilde{t}\,), \wit{\xi}(\tilde{t}'\,)\bigr]\bigr\rangle 
= 2\II \hbar\tilde{\gamma}\frac{\D}{\D \tilde{t}}\delta\bigl(\tilde{t}-\tilde{t}'\bigr)
\EEQ
\end{subequations}
where we also re-scaled $\tilde{T} := T/\lambda$ and the distribution $I$ has a known integral representation 
(see \cite{Gard04,Weis21,Arau19,Wald21} and (\ref{3.8a}) below). The reduced description (\ref{3.6}) does not apply to
the early-time regime, but since the properties of that regime are non-universal anyway, one cannot hope to study them
through the perspective of extremely simplified models such as considered here. 

Dropping the tildes throughout and focussing on the leading low-momentum behaviour, 
we have found that the long-time behaviour of the quantum Langevin equations (\ref{3.2},\ref{3.3}) 
simply follows from the over-damped Langevin equation ($k=|\Vek{k}|$) 
\BEQ \label{3.7}
\gamma \partial_t \phi_{\Vek{k}}(t) + \bigl( r(t) + k^2 \bigr) \phi_{\Vek{k}}(t) = \xi_{\Vek{k}}(t)
\EEQ
and that the physical nature of the dynamics will be determined by the form of the noise correlators involving the $\xi_{\Vek{k}}(t)$. 
In addition, since we aim at an understanding of the relevance of quantum noise for the long-time dynamics, we shall from now on
concentrate on the zero-temperature limit\footnote{If $T>0$, expect after a finite time 
loss of quantum coherence and hence classical dynamics.} and let $T\to 0$. A major qualitative difference between
quantum and classical dynamics is the non-markovianity of quantum noise. Another important difference comes from different scalings. 
In order to elucidate their respective relevance, we shall consider the following three types of noise correlators \cite{Wald21}:
\begin{enumerate}
\item {\em quantum noise}, at temperature $T=0$ and a regulator $\mathfrak{t}_0\sim 1/\gamma$, is defined by 
\begin{subequations}\label{3.8}\begin{align}
\bigl\langle\bigl\{ \xi_{\Vek{k}}(t), \xi_{\Vek{k}}(t)\bigr\}\bigr\rangle &=
\frac{\gamma\hbar}{\pi} \int_{\mathbb{R}}\!\D\omega\: |\omega|e^{\II\omega(t-t')} e^{-\mathfrak{t}_0|\omega|}\,\delta(\Vek{k}+\Vek{k}') 
\nonumber \\
&= \frac{\gamma\hbar}{\pi} \frac{\mathfrak{t}_0^2 -(t-t')^2}{[\mathfrak{t}_0^2+(t-t')^2]^2}\,\delta(\Vek{k}+\Vek{k}') \label{3.8a}\\
\bigl\langle\bigl[ \xi_{\Vek{k}}(t), \xi_{\Vek{k}}(t)\bigr]\bigr\rangle &=
2\II\hbar\gamma \left( \frac{\D}{\D t}\delta(t-t')\right)\delta(\Vek{k}+\Vek{k}') \label{3.8b}
\end{align}\end{subequations}
These noise correlators follow from the quantum correlators (\ref{3.3},\ref{3.6b}). The regularisation merely permits a finite expression of
the associated distribution. Non-universal quantities such as critical exponents should turn out to be independent of 
$\mathfrak{t}_0$, whereas non-universal quantities such as the location of the critical point may depend on it. At the end, one should
strive at taking $\mathfrak{t}_0\to 0$, consistently with the over-damped limit $\gamma\to \infty$ implicit in the over-damped
equation (\ref{3.7}). 
\item {\em effective noise} is defined by the correlators
\begin{align} \label{3.9}
\bigl\langle\bigl\{ \xi_{\Vek{k}}(t), \xi_{\Vek{k}}(t)\bigr\}\bigr\rangle &=
\mu |\Vek{k}|^2 \delta(t-t')\,\delta(\Vek{k}+\Vek{k}') \;\; ~;~ \;\;
\bigl\langle\bigl[ \xi_{\Vek{k}}(t), \xi_{\Vek{k}}(t)\bigr]\bigr\rangle = 0 
\end{align}
Herein, the scaling $\sim \frac{1}{(t-t')^2}$ (in the $\mathfrak{t}_0\to 0$ limit) is replaced by an equivalent scaling in the momentum $|\Vek{k}|$
and $\mu$ serves as a control parameter.  
Hence quantum and effective noises have the same scaling properties, but effective noise is markovian while quantum noise is not. Comparison of the
results of both will permit to appreciate the importance of non-markovian effects for the long-time dynamics. 
\item {\em classical white noise} is of course defined by 
\begin{align} \label{3.10}
\bigl\langle\bigl\{ \xi_{\Vek{k}}(t), \xi_{\Vek{k}}(t)\bigr\}\bigr\rangle &=
4\gamma\, T\, \delta(t-t')\,\delta(\Vek{k}+\Vek{k}') \;\; ~;~ \;\;
\bigl\langle\bigl[ \xi_{\Vek{k}}(t), \xi_{\Vek{k}}(t)\bigr]\bigr\rangle = 0 
\end{align}
and differs in its scaling from effective noise. It is clearly markovian. 
\end{enumerate} 
In what follows, we shall compare the behaviour of three distinct types of dynamics:
\begin{enumerate}
\item {\em quantum dynamics}, given by (\ref{3.7},\ref{3.8}). 
\item {\em effective dynamics}, given by (\ref{3.7},\ref{3.9}). 
\item {\em classical dynamics}, given by (\ref{3.7},\ref{3.10}). 
\end{enumerate}
The results will be interpreted in the context of physical ageing. 

\section{What is physical ageing~?}

From the solution $\phi_{\Vek{k}}(t)$ of the equation of motion (\ref{3.7}), we define the observables:
\begin{enumerate}
\item {\em equal-time correlations} $C_{\Vek{k}}(t)$ are obtained as
\BEQ \label{4.1}
\delta(\Vek{k}+\Vek{k}')\, C_{\Vek{k}}(t) := \bigl\langle \bigl\{ \phi_{\Vek{k}}(t),\phi_{\Vek{k}'}(t)\bigr\}\bigr\rangle
\EEQ
to be studied in the long-time scaling limit (hence the dynamical exponent $z=2$)
\BEQ \label{4.2}
t\to\infty\;\; , \;\; k = |\Vek{k}|\to 0 \;\; , \;\; \mbox{\rm ~such that $\rho:= k^2 t/\gamma$ is kept fixed}
\EEQ
\item {\em two-time correlations} $C_{\Vek{k}}(t,s)$ are obtained as
\BEQ \label{4.3}
\delta(\Vek{k}+\Vek{k}')\, C_{\Vek{k}}(t,s) := \bigl\langle \bigl\{ \phi_{\Vek{k}}(t),\phi_{\Vek{k}'}(s)\bigr\}\bigr\rangle
\EEQ
The {\em auto-correlator} is $C(t,s) := \int_{\Vek{k},(\Lambda)}C_{\Vek{k}}(t,s)$, denoting  
$\int_{\Vek{k},(\Lambda)} :=  \int_0^{\Lambda} \int_{S^d} \frac{\D\Vek{k}}{(2\pi)^d}$ such that $\Lambda$ describes an {\sc uv}-cutoff. 
Any quantity depending explicitly on $\Lambda$ cannot be universal. 
\item {\em two-time responses} $R_{\Vek{k}}(t,s)$ are obtained 
as
\BEQ \label{4.4}
R_{\Vek{k}}(t,s) := \left.\frac{\delta \bigl\langle \phi_{\Vek{k}}(t)\bigr\rangle}{\delta h_{\Vek{k}}(s)}\right|_{h=0}
\EEQ
where $h$ is the magnetic field conjugate to the magnetisation $\bigl\langle \phi_{\Vek{k}}(t)\bigr\rangle$. The {\em auto-response}
is $R(t,s) := \int_{\Vek{k},(\Lambda)}R_{\Vek{k}}(t,s)$. Herein, $t$ is called the {\em observation time} and $s$ the {\em waiting time}.
\end{enumerate}

{\it Physical ageing} was originally observed in the slow dynamics of glasses after a quench from a melt to below the glass-transition temperature \cite{Struik78}. 
Here, we shall characterise the initial state through a vanishing magnetisation and the initial equal-time correlator
\BEQ \label{4.5}
\bigl\langle \phi_{\Vek{k}}(0)\bigr\rangle = 0 \;\; ~~;~~ C_{\Vek{k}}(0) = c_0 + c_{\alpha} |\Vek{k}|^{\alpha}
\EEQ
which in direct space means $C(0,\Vek{x})\sim |\Vek{x}|^{-d-\alpha}$ such that for $\alpha\geq 0$ the initial correlations are short-ranged and
for $\alpha<0$ they are long-ranged. Numerous studies in classical systems lead to the following expectations, see \cite{Godr00b,Cugl03,Henkel10}:
\begin{enumerate}
\item for a quench into the disordered phase $T>T_c$ (or $\lambda>\lambda_c$) the systems rapidly become time-translation-invariant\footnote{Please do not confuse
time-translation-invariant two-time correlators $C_{\Vek{k}}(\tau)$ and responses $R_{\Vek{k}}(\tau)$ from (\ref{4.6}) with an equal-time correlator $C_{\Vek{k}}(t) := C_{\Vek{k}}(t,t)$.} 
\BEQ \label{4.6}
C_{\Vek{k}}(t,s) = C_{\Vek{k}}(t-s) = C_{\Vek{k}}(\tau) \;\; ~~;~~
R_{\Vek{k}}(t,s) = R_{\Vek{k}}(t-s) = R_{\Vek{k}}(\tau)
\EEQ
and $C_{\Vek{k}}(\tau)$ and $R_{\Vek{k}}(\tau)$ should decay exponentially fast with $\tau$. Since there is a single stationary (equilibrium) state,
the (quantum) fluctuation-dissipation theorem should hold but no ageing is expected. 
\item for a quench onto criticality or into the ordered phase $T\leq T_c$ (or $\lambda\leq \lambda_c$), there is no time-translation-invariance. 
If the dynamics can be described in terms of a single length scale $L(t)\sim t^{1/z}$, one finds dynamical scaling for 
$t\gg \tau_{\rm micro}$, $s\gg \tau_{\rm micro}$ and $t-s\gg \tau_{\rm micro}$  ($\tau_{\rm mirco}$ is a microscopic reference time-scale)
\BEQ \label{4.7}
C(t,s) = s^{-b} f_C\left(\frac{t}{s}\right) \;\; ~~;~~ \;\; R(t,s) = s^{-1-a} f_R\left(\frac{t}{s}\right)
\EEQ
with the asymptotic behaviour $f_{C,R}(y)\sim y^{-\lambda_{C,R}/z}$ with 
defines\footnote{Please do not confuse the autocorrelation exponent $\lambda_C$ with the critical point $\lambda_c$.}  
the {\em autocorrelation exponent} $\lambda_C$ and the {\em auto-response exponent} $\lambda_R$.
The exponents $a,b$ are called {\em ageing exponents}. Strong fluctuations lead to a breaking of the fluctuation-dissipation theorem.  
If this holds true, the three defining properties of {\em physical ageing} \cite{Henkel10}, 
namely (i) slow dynamics, (ii) breaking of time-translation-invariance and (iii) dynamical scaling are satisfied. 
\end{enumerate}
In the classical spherical model, these expectations are fully borne out \cite{Ronc78,Cugl94,Godr00b}. For the initial conditions
(\ref{4.5}), the exact values of all exponents are known for the spherical model \cite{Pico02}, see table~\ref{tab2} below. 
For detailed reviews, see \cite{Cugl03,Henkel10}. 

Are these classically motivated expectations verified in quenched quantum dynamics~? It is often thought that the answer should be
affirmative: {\it ``. . . a large class of coarsening systems (classical, quantum, pure and disordered)
should be characterised by the same scaling functions.''} \cite{Aron09}. Is this always so~? In what follows, we shall study this question
for the quantum spherical model, at temperature $T=0$. 

\section{The spherical constraint}

Having brought together in sections~1-4 the physical background for studying non-equilibrium quantum dynamics and ageing, we now turn to the
exact solution of the quantum spherical model at $T=0$ and describe the results \cite{Wald21}. 

The formal solution of the equation of motion (\ref{3.7}) is
\BEQ \label{5.1}
\phi_{\Vek{k}}(t) = \frac{\exp\bigl(-k^2 t/\gamma\bigr)}{\sqrt{g(t)\,}} 
\left[ \phi_{\Vek{k}}(0) + \frac{1}{\gamma} \int_0^t \!\D t'\: \sqrt{g(t')\,}\, \exp\bigl( k^2t'/\gamma\bigr)\xi_{\Vek{k}}(t') \right]
\EEQ
with the important auxiliary function
$g(t) := \exp\left( \frac{2}{\gamma} \int_0^t \!\D t'\: r(t') \right)$. 
In order to re-write the spherical constraint (\ref{2.2}) as an equation for
$g(t)$, we first define two further supplementary functions\\[-0.6truecm]
\begin{subequations}\begin{align}
A(t) &= c_{\alpha} A_{\alpha}(t) := \int_{\mathbb{R}^d} \!\D\Vek{k}\: \exp\left(-2\frac{k^2 t}{\gamma}\right) c_{\alpha} k^{\alpha} \\
F(t,s) &:= \int_{\mathbb{R}^d} \!\D\Vek{k}\: \exp\left(-\frac{k^2 (t+s)}{\gamma}\right)
\bigl\langle \bigl\{ \xi_{\Vek{k}}(t), \xi_{-\Vek{k}}(s)\bigr\}\bigr\rangle
\end{align}\end{subequations}
and also $g_2(t,s) := \sqrt{g(t) g(s)\,}$. The spherical constraint (\ref{2.2}) then becomes, using the definition (\ref{4.1}) and
the solution (\ref{5.1}) 
\BEQ
\frac{1}{\lambda} \stackrel{!}{=} C(t,t) = \frac{1}{g(t)} \left[ A(t) + \bigl( g_2 * * F\bigr)(t,t) \right]
\EEQ
where $\bigl( h_1 ** h_2\bigr)(t,s) := \int_0^t \!\D x\int_0^s \!\D y\: h_1(x,y) h_2(t-x,s-y)$
is the two-dimensional convolution. It follows that the spherical constraint fixes the function $g(t)$
\BEQ \label{5.5}
\frac{1}{\lambda} g(t) = A(t) + \bigl( g_2 * * F\bigr)(t,t) 
\EEQ

\subsection{Markovian case}
If the noise correlator $\bigl\langle\bigl\{ \xi_{\Vek{k}}(t), \xi_{\Vek{k}}(t)\bigr\}\bigr\rangle\sim \delta(t-t')$ is markovian, 
the constraint (\ref{5.5}) turns into a linear Volterra equation (here for effective noise)
\BEQ
\frac{1}{\lambda} g(t) = A(t) + \frac{\mu}{\gamma^2} \bigl( g *  A_2\bigr)(t) 
~~\Longrightarrow~~ \lap{g}(p) = \frac{c_{\alpha}\lap{A}_{\alpha}(p)}{1/\lambda- \mu/\gamma^2\, \lap{A}_2(p)}
\EEQ
which is formally solved by a Laplace transformation $\lap{h}(p) = \int_0^{\infty}\!\D t\: e^{-pt} h(t)$. 
The remainder of the procedure is now standard. Tauberian theorems \cite{Fell71} state that the long-time behaviour of $g(t)$ for $t\to\infty$ 
is related to the one of $\lap{g}(p)$ for $p\to 0$. The critical point $\lambda_c$ is given by the smallest p\^ole of $\lap{g}(p)$. 
Expanding $\lap{A}_{\alpha}(p)$ around $p=0$ it follows that for quenches to $\lambda>\lambda_c$, one has $g(t)\sim e^{t/\tau_r}$ and
for quenches to $\lambda\leq \lambda_c$, one finds $g(t)\sim t^{\digamma}$, with the values of $\digamma$ listed in table~\ref{tab1} below. 

\subsection{Non-markovian case}
In the non-markovian case, (\ref{5.5}) is a non-linear integral equation for $g(t)$. Progress can be made by considering instead the
symmetric function $G(t,s)=G(s,t)$ which satisfies the equation
\BEQ \label{5.7}
\frac{1}{\lambda} G(t,s) = A\left( \frac{t+s}{2} \right) + \bigl( G** F\bigr)(t,s)
\EEQ
which reduces to (\ref{5.5}) in the limit $s\to t$, hence $g(t)=G(t,t)$ (although $G(t,s)\ne g_2(t,s)$). 
Denote by $\dlap{h}(p,q)=\int_0^{\infty}\!\D x\int_0^{\infty}\!\D y\: e^{-px-qy} h(x,y)$ the two-dimensional
Laplace transform. Then the formal solution of (\ref{5.7}) is
\BEQ
\dlap{G}(p,q) = \frac{\dlap{A}(p,q)}{1/\lambda - \dlap{F}(p,q)}
\EEQ
The interpretation of this result is again via a Tauberian theorem.\\

\noindent \begin{subequations}{\bf Lemma}: \cite{Wald21} {\it For a homogeneous function $f(x,y)=y^{-\alpha}\phi(x/y)$ with $\phi(0)$ finite and 
asymptotically $\phi(u)\stackrel{u\gg1}{\simeq}\phi_{\infty}\, u^{-\lambda}$, one has the scaling form}
\begin{align}
\dlap{f}(p,q) = p^{\alpha-2} \Phi(q/p) \;\; , \;\; 
\Phi(u) = \Gamma(2-\alpha) u^{\alpha-1}\int_0^{\infty} \!\D\xi\: \phi(\xi u)(\xi+1)^{\alpha-2}
\end{align}
{\it If $n<\lambda<n+1$ with $n\in\mathbb{N}$, one has asymptotically for $u\to\infty$}
\begin{align}
\Phi(u) &\simeq \phi^{(1)} u^{\alpha-2} + \ldots + \phi^{(n)} u^{\alpha-1-n} + \Phi_{\infty}\, u^{\alpha-1-\lambda} \\
\Phi_{\infty} &= \phi_{\infty} \frac{\Gamma(1-\lambda)}{\Gamma(1+\lambda-\alpha)} \;\; , \;\;
\phi^{(m)} = (-1)^{m-1} \frac{\Gamma(m+1-\alpha)}{(m-1)!} \int_0^{\infty} \!\D u\: u^{m-1} \phi(u) \nonumber
\end{align}\end{subequations}

The critical point is found from the smallest p\^ole of $\dlap{G}(p,q)$. Expanding $\dlap{F}(p,q)$, this gives $\frac{1}{\lambda_c}=\dlap{F}(0,0)$. 
Explicitly ($\Omega_d=|S^d|$, $C_E=0.5772\ldots$ is Euler's constant) 
\BEQ
\frac{1}{\lambda_c} = \left\{
\begin{array}{ll} \frac{\mu}{\gamma} \frac{\Omega_d}{(2\pi)^d}\frac{\Lambda^d}{d}  & \mbox{\rm ~~;~ effective noise}  \\
-\frac{4\hbar}{\pi\gamma}\frac{\Omega_d}{(2\pi)^d} \left\{
\frac{\Lambda^d}{d} \left[ \ln\left(\Lambda^2\frac{\mathfrak{t}_0}{\gamma}\right)+C_E -\frac{2}{d}\right] + {\rm O}(\mathfrak{t}_0) \right\} 
& \mbox{\rm ~~;~ quantum noise}  \end{array}\right.
\EEQ
which is finite for all $d>0$. Hence $d_{\ell}=0$ for both quantum dynamics and effective dynamics which is different from the equilibrium values of 
$d_{\ell}$ quoted in (\ref{2.3}). Hence {\em the stationary state of the $T=0$ quantum dynamics cannot be an equilibrium state}~! This
is even more surprising since the single-particle dynamics constructed in section~1 should for any $T>0$ relax to the unique equilibrium state. 

Qualitatively, the results for $g(t)$ of non-markovian quantum dynamics are analogous to the ones of effective dynamics. For quenches to
$\lambda>\lambda_c$, $g(t)\sim e^{t/\tau_r}$ is exponential, with $\tau_r \sim \bigl( \lambda - \lambda_c\bigr)^{-2/d}$. 
For quenches to $\lambda\leq \lambda_c$, we read off 
$\dlap{G}(p,q) = p^{-\digamma-2}\mathbb{G}(q/p)$, hence $G(t,s)=s^{\digamma}\mathscr{G}(t/s)$ by the Lemma. It follows that
$g(t) = G(t,t)= t^{\digamma}\mathscr{G}(1)$ and the values of $\digamma$ are listed in table~\ref{tab1}. 
They are the same as for effective dynamics. The constant $\mathscr{G}(1)$ will not be needed in the leading terms of the observables. 

\begin{table}
\begin{tabular}{|c|lc||c|c|c|c|c|}\hline
\multicolumn{2}{|c||}{quantum region}  & $\digamma$ & $\lambda_C$ & $\lambda_R$ & $a$ & $b$  \\ \hline \hline
\multirow{3}{*}{{$\lambda=\lambda_c$}} & 
\multicolumn{1}{|l||}{{\bf I}~~ $0<d<2$}              & $-\frac{\alpha}{2}$    & $d+\frac{\alpha}{2}$   & $d-\frac{\alpha}{2}$   & $\frac{d}{2}-1$ & $\frac{d}{2}$ \\ \cline{2-7}
&\multicolumn{1}{|l||}{{\bf II}~ $2<d$, $d+\alpha<2$} & $1-\frac{d+\alpha}{2}$ & $1+\frac{d+\alpha}{2}$ & $\frac{d-\alpha}{2}+1$ & $\frac{d}{2}-1$ & $1$ \\ \cline{2-7}
&\multicolumn{1}{|l||}{{\bf V}~ $2<d$, $d+\alpha>2$}  & 0                      & $d+\alpha$       	    & $d$              	     & $\frac{d}{2}-1$ & $\frac{d+\alpha}{2}$ \\ \hline 
\multicolumn{2}{|l||}{{$\lambda<\lambda_c$}}          & $-\frac{d+\alpha}{2}$  & $\frac{d+\alpha}{2}$   & $\frac{d-\alpha}{2}$   & $\frac{d}{2}-1$ & $ 0$ \\ \hline
\end{tabular}
\caption[tab1]{Non-equilibrium exponents of the quantum spherical model for $\lambda\leq \lambda_c$ at $T=0$ \cite{Wald21}.\label{tab1}} 
\end{table}

\section{Quench into the dis-ordered phase}

We now review results for a quantum quench with $\lambda>\lambda_c$ \cite{Wald21}, where $g(t)\sim e^{t/\tau_r}$.  
For the stationary single-time correlator $C_{\Vek{k}}(\infty)$, we find 
\BEQ
C_{\Vek{k}}(\infty) \simeq \left\{
\begin{array}{ll} \frac{\mu}{\gamma^2}\frac{k^2}{1/\tau_{\text{r}} + 2 k^2/\gamma} & \mbox{\rm ~~;~ effective noise} \\
\frac{\hbar}{\pi \gamma} g_{\rm AS}\left( \mathfrak{t}_0 
\left( k^2/\gamma + (2\tau_{\text{r}})^{-1} \right) \right) & \mbox{\rm ~~;~ quantum noise}
\end{array}\right.
\EEQ
where $g_{\rm AS}(x) := \int_0^\infty  \!\!\D t\, \frac{\cos t}{t+x}\simeq x^{-2}$ for $x\gg 1$. They are quite distinct,
but the result of quantum noise is qualitatively very similar to the classical Ornstein-Zernicke form. 

Next, the two-time correlators $C_{\Vek{k}}(s+\tau,s)$ do indeed satisfy time-translation-invariance for $s\gg \tau_{\rm micro}$, as expected
\BEQ \label{6.2}
C_{\Vek{k}}(s+\tau,s) \simeq \left\{ \begin{array}{ll}
\frac{\mu k^2}{\gamma^2} 
\frac{1}{\frac{1}{\tau_{\text{r}}}+2\frac{k^2}{\gamma}} 
\exp \left(-\left(\frac{1}{2\tau_{\text{r}}} +\frac{k^2}{\gamma}\right)\tau \right) & \mbox{\rm ~~;~ effective noise} \\
-\frac{2\hbar}{\pi \gamma} 
\frac{1}{\left[(2\tau_{\text{r}})^{-1} + k^2/\gamma \right]^2} 
\frac{1}{\tau^2} & \mbox{\rm ~~;~ quantum noise}
\end{array} \right.
\EEQ
but their functional forms are very different. Analogously, the two-time response $R_{\Vek{k}}(s+\tau,s)$ is time-translation-invariant 
for $s\gg \tau_{\rm micro}$, with the same form for both effective and quantum noises\\[-0.2truecm] 
\BEQ \label{6.3}
R_{\Vek{k}}(s+\tau,s) \simeq 
\frac{1}{\gamma} \exp\left(-\left(\frac{1}{2\tau_{\text{r}}}+\frac{k^2}{\gamma}\right)\tau\right)
\EEQ
For effective noise, correlators and responses decay exponentially with $\tau$. Empirically, one might say that they satisfy
an `effective fluctuation-dissipation theorem'\footnote{Here $C_{\Vek{k}}(\tau) := C_{\Vek{k}}(s+\tau,s)$ and $R_{\Vek{k}}(\tau) := R_{\Vek{k}}(s+\tau,s)$ denote two-time
correlators and responses, respectively, which have become independent of the waiting time $s$, if $s$ is large enough.} 
\BD
\frac{\partial C_{\Vek{k}}(\tau)}{\partial \tau} = -\frac{\mu}{2\gamma} \bigl|\Vek{k}\bigr|^2 R_{\Vek{k}}(\tau) 
= - \frac{T_{\rm eff}(k)}{\gamma} R_{\Vek{k}}(\tau) 
\ED
but therein $T_{\rm eff}(k)$ is distinct from the bath temperature $T=0$. So that relation is rather {\it ad hoc}. 
For quantum noise, the different forms of $C_{\Vek{k}}(\tau)$ and $R_{\Vek{k}}(\tau)$ in (\ref{6.2},\ref{6.3}) exclude the validity of any fluctuation-dissipation theorem. 

In conclusion, although there is a single stationary state of the dynamics, {\em this stationary state cannot be an equilibrium state,} neither
for effective nor for quantum noise, since the {\sc qfdt} does not hold.  

\section{Quench onto criticality or into the ordered phase}

We now review results for a quantum quench with $\lambda\leq\lambda_c$ \cite{Wald21}, where throughout $g(t)\sim t^{\digamma}$. 
For a quenched into the ordered phase
with $\lambda<\lambda_c$, the dynamics is the same as in the classical case and only depends on the initial correlations (\ref{4.5}). 
This is expected, since the same already occurs for classical white noise dynamics \cite{Pico02} 
and quantum noise is more weak than classical noise, see figure~\ref{fig2}. 

\begin{figure}[tb]
\includegraphics[width=.95\hsize]{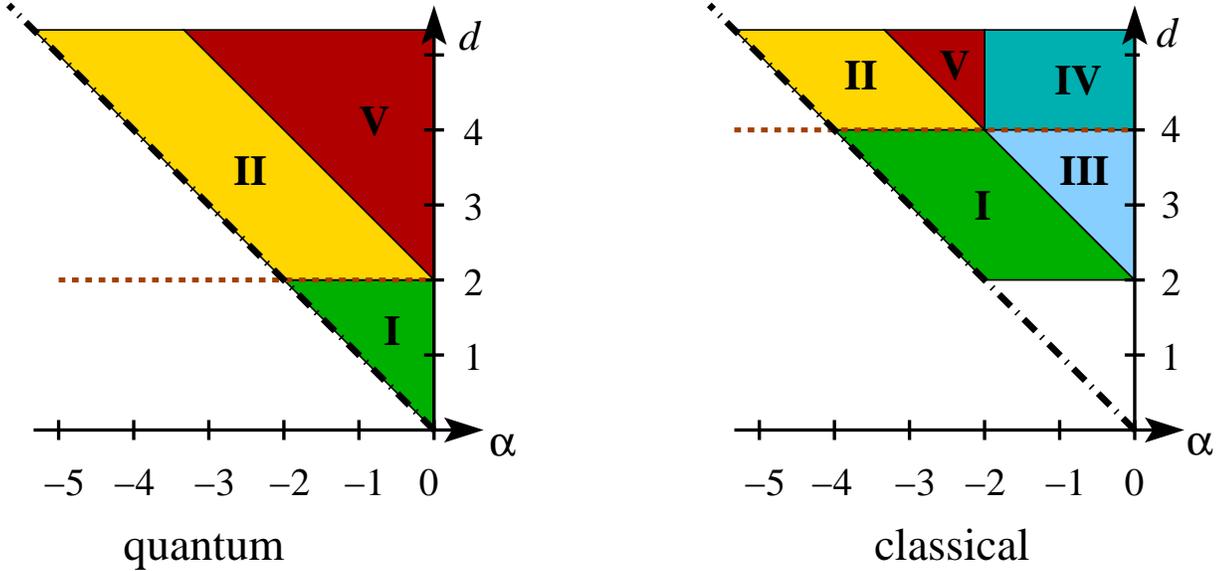} 
\caption[fig5]{Critical scaling regions for quantum and classical dynamics. \label{fig5}}
\end{figure}
\begin{figure}[t]
\includegraphics[width=.95\hsize]{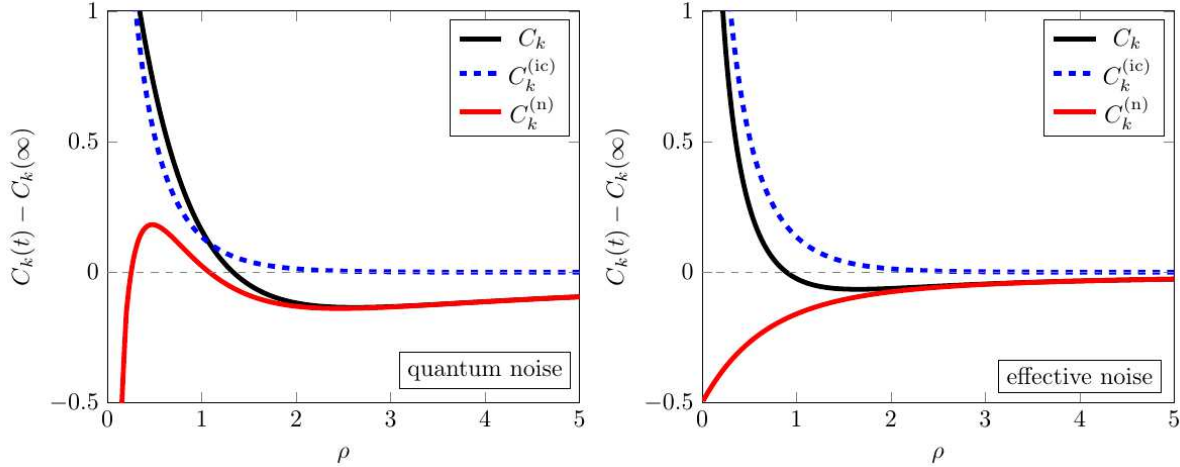} 
\caption[fig6]{Critical equal-time correlator $C_{\Vek{k}}(t)-C_{\Vek{k}}(\infty)$ as a function of $\rho = k^2 t/\gamma$ in region {\bf I}. \label{fig6}}
\end{figure}

For critical quenches to $\lambda=\lambda_c$, one obtains
several scaling regions, as shown in figure~\ref{fig5} in dependence of the values of the dimension $d$ and the parameter $\alpha$ of the
initial correlations (\ref{4.5}). The regions are the same for quantum and effective dynamics. 
The dotted horizontal line indicates the upper critical dimension $d_u$, and we read off\footnote{In agreement with the results of Keldysch field-theory of the O($\infty$) model \cite{Gagel14,Gagel15}.} 
$d_u^{({\rm qu})}=2$ and $d_u^{({\rm cl})}=4$. In the quantum case, this is different from the equilibrium values of $d_u$ quoted in (\ref{2.3}). The regions are characterised as follows:
\begin{enumerate}
\item[{\bf I.}] both bath and initial fluctuations are relevant. 
\item[{\bf II.}] only initial long-ranged fluctuations are relevant. 
\item[{\bf V.}] no relevant fluctuations at all, long-ranged initial correlations. 
\end{enumerate}
For classical dynamics, two more regions exist, without a quantum counterpart (for short-ranged initial correlations with $\alpha=0$ these
are the main cases for study):
\begin{enumerate}
\item[{\bf III.}] thermal bath fluctuations are relevant. 
\item[{\bf IV.}] no relevant fluctuations at all, short-ranged initial correlations. 
\end{enumerate}
The exact values of the non-equilibrium exponents are listed in table~\ref{tab1} for quantum and effective dynamics and in table~\ref{tab2}
for classical dynamics. We see that in region {\bf I} there is a shift $d\mapsto d-1$ in $\lambda_C$ and $\lambda_R$ when going from quantum to 
classical. Otherwise, the exponents are identical (the admissible values of $d$ and $\alpha$ can be different). 
Although the exponents are the same for quantum and effective dynamics, the scaling functions can be different. This is shown in fig.~\ref{fig6} 
\cite{Wald21} for the equal-time correlator $C_{\Vek{k}}(t)=C_{\Vek{k}}^{(ic)}(t)+C_{\Vek{k}}^{(n)}(t)$ in region {\bf I}. The contributions
$C_{\Vek{k}}^{(n)}(t)$ of quantum and effective noise are different, while the initial contribution $C_{\Vek{k}}^{(ic)}(t)$ obviously is the same. 
In regions {\bf II} and {\bf V}, only $C_{\Vek{k}}^{(ic)}(t)$ is relevant. All scaling functions are known
analytically \cite{Wald21}. Analogous statements hold true for
the two-time correlator $C_{\Vek{k}}(t,s)$ while the form of the two-response $R_{\Vek{k}}(t,s)$ is noise-independent. 

\begin{table}[b]
\begin{tabular}{|l|lr||c||c|c|c|c|c|}\hline
\multicolumn{3}{|c||}{classical region}       & $\digamma$             & $\lambda_C$            & $\lambda_R$      	     & $a$ 		       & $b$    \\ \hline 
\multirow{5}{*}{{$T=T_c$}} & 
{\bf I}$_c$    & {\fns$2<d<4$, $0<d+\alpha<2$} & $-1-\frac{\alpha}{2}$  & $d+\frac{\alpha}{2}-1$ & $d-\frac{\alpha}{2}-1$ & $\frac{d}{2}-1$ & $\frac{d}{2}-1$ \\ \cline{2-8}
&{\bf II}$_c$  & {\fns$4<d$, $0<d+\alpha<2$}   & $1-\frac{d+\alpha}{2}$ & $1+\frac{d+\alpha}{2}$ & $\frac{d-\alpha}{2}+1$ & $\frac{d}{2}-1$ & $1$    \\ \cline{2-8}
&{\bf III}$_c$ & {\fns$2<d<4$, $d+\alpha>2$}   & $\frac{d}{2}-2$        & $\frac{3}{2}d-2$ & $\frac{3}{2}d-2$  & $\frac{d}{2}-1$ & $\frac{d}{2}-1$   \\  \cline{2-8}
&{\bf IV}$_c$  & {\fns $4<d$, $d+\alpha>2$, $\alpha>-2$} & 0            & $d$                    & $d$                    & $\frac{d}{2}-1$ & $\frac{d}{2}-1$ \\  \cline{2-8}
&{\bf V}$_c$   & {\fns$4<d$, $d+\alpha>2$, $\alpha<-2$}  & 0            & $d+\alpha$             & $d$                    & $\frac{d}{2}-1$ & $\frac{d+\alpha}{2}$\\ \hline 
\multicolumn{2}{|l}{{$T<T_c$}}    & $2<d$      & $-\frac{d+\alpha}{2}$  & $\frac{d+\alpha}{2}$   & $\frac{d-\alpha}{2}$   & $\frac{d}{2}-1$ & $ 0$ \\ \hline
\end{tabular}
\caption[tab2]{Non-equilibrium exponents of the classical spherical model for $T\leq T_c$ \cite{Pico02}.\label{tab2}} 
\end{table}

\section{Conclusions}

Several surprises arise in the $T=0$ quantum dynamics of the spherical model:
\begin{enumerate}
\item the stationary state is not a quantum equilibrium state , not even for $\lambda>\lambda_c$
\item the non-equilibrium exponents for $\lambda\leq\lambda_c$ are insensitive to non-markovianity
\item non-markovian noise is important for equal-time correlators 
\end{enumerate}
Figure~\ref{fig5} shows the correspondence of the critical scaling regimes for quantum and classical dynamics. 
The qualitative scenario of physical ageing (section~4) is confirmed, but a comparison of tables~\ref{tab1} and~\ref{tab2} shows that the
values of the exponents are different. 

Turning to possible dynamical symmetries, the underlying equation (\ref{3.7}) has $z=2$ and does
admit a dynamical Schr\"odinger symmetry if $r(t)\sim t^{-1}$ \cite{Pico04,Henkel10,Stoi22}.  
This ansatz for $r(t)$ does hold true for classical dynamics at $T\leq T_c$. For quantum dynamics at $T=0$ and $\lambda\leq\lambda_c$, 
$r(t)=\frac{\gamma}{2}\partial_t \ln g(t)\simeq\frac{\gamma}{2}\frac{\digamma}{t}$. In region {\bf I}, since $\digamma=-\frac{\alpha}{2}\to 0$
in the limit of short-ranged initial conditions $\alpha=0$, the ansatz\footnote{It is also used in Keldysch field theory with disordered initial state \cite{Gagel15}.} 
$r(t)\sim t^{-1}$ no longer applies. New representations for a dynamical symmetry must be sought. 


\noindent
{\bf Acknowledgements}\\
It is a pleasure to thank R. Ara\'ujo, A. Gambassi, M. Hase, G.T. Landi  and especially and foremost S. Wald for year-long exchanges and collaborations 
on the quantum dynamics of the spherical model. Financial support by PHC Rila (KP-06-RILA/7) is gratefully acknowledged.

\end{document}